\documentclass[pop]{revtex4}
\usepackage{graphicx}
\usepackage{amsmath}

\usepackage[utf8]{inputenc}
\usepackage{amssymb}
\usepackage{dsfont}
\usepackage{gensymb}
\usepackage{makecell}
\allowdisplaybreaks

\DeclareUnicodeCharacter{00EF}{i}
\newlength{\myfiguerwidth}

\begin{document}

\title{A method to achieve rapid localised deep heating in a laser irradiated solid density target}

\author{H.~Schmitz}
\email{holger.schmitz@stfc.ac.uk}

\author{A.P.L.~Robinson}

\affiliation{Central Laser Facility, STFC, Rutherford Appleton Laboratory, Didcot,
Oxon., OX11 0QX}

\begin{abstract}
Rapid heating of small buried regions by laser generated fast electrons may be useful for applications such as XUV radiation sources or as drivers for shock experiments. In non-structured targets the heating profile possesses a global maximum near the front surface. This paper presents a new target design that uses resistive guiding to concentrate the fast electron current density at a finite depth inside the target. The choice of geometry uses principles of non-imaging optics. A global temperature maximum at depths up to $50\mu$m into the target is achieved. Although theoretical calculations suggest that small source sizes should perform better than large ones, simulations show that a large angular spread at high intensities results in significant losses of the fast electrons to the sides. A systematic parameter scan suggests an optimal laser intensity. A ratio of 1.6 is demonstrated between the maximum ion temperature and the ion temperature at the front surface.
\end{abstract}


\maketitle

\setlength{\myfiguerwidth}{\columnwidth}

\section{Introduction \label{SecIntroduction}}

When a high intensity laser impinges on a solid density target a large part of the electromagnetic energy of the laser is converted into kinetic energy of fast moving electrons. The ponderomotive force of the laser accelerates electrons to relativistic energies and typically around 15\% to 30\% of the laser energy is converted \cite{Nakatsutsumi:2008} while conversion rates up to 90\% have been shown \cite{Ping:2008}. These electrons can be useful for a number of different applications such as the Fast Ignition scheme for inertial confinement fusion \cite{Tabak:1994}, ion acceleration \cite{Wilks:2001}, or as a means to heat localised regions within the target \cite{Robinson:2013}. In the last of these applications the local hot spots can act as XUV radiation sources \cite{Park:2006}, as drivers for shock experiments \cite{Robinson:2013} or as drivers of jets in laboratory astrophysics experiments \cite{Schmitz:2015}. In all cases there is a need to control the propagation of the fast electrons inside the solid density target.

One method of influencing the propagation of fast electrons is through magnetic fields, self generated at resistivity gradients \cite{Robinson:2007,Robinson:2012}. Using a number of materials with different Z to construct the target can generate multi kilotesla magnetic fields at the material interfaces \cite{Robinson:2007,Ramakrishna:2010}. These magnetic fields are strongly localised to a very narrow region around the interface, on a scale length comparable to the electron skin depth. These thin magnetic field layers reflect electrons that intercept the layer at sufficiently oblique angles. It has been shown that, for reasons of symmetry, the reflection is specular \cite{Schmitz:2012a}. Because the motion of the electrons in the homogeneous domains can be assumed to be ballistic, one can apply the concepts of geometric mirror optics to the design of targets that guide and focus fast electrons.

The aim of this paper is to investigate possible geometries for heating a small, well defined region within a solid density target. To this end, the fast electrons that originate from the laser-plasma interaction region near the critical density surface need to be focused into a small volume. The fast electron current in this volume will draw a neutralising return current of cold background electrons in order to preserve global charge neutrality. This return current is responsible for depositing the fast electron energy into the target via Ohmic heating. It is therefore the aim to concentrate the electrons as much as possible to create an optimal hot spot inside the target.

In order to understand the limitations of this endeavour, section \ref{SecEtendue} will review the optical concept of etendue. The etendue is a conservation property that is linked to the Liouville equation and provides a quantitative measure of how well a diffuse source can be collimated. It is linked to the source radius and its angular spread. In section \ref{SecSimulations} we will investigate the ability of specially designed targets to collimate electrons into a small region deep inside the target. The design of the targets is inspired by the concepts of non-imaging optics, specifically it will be a variant of a Compound Elliptical Concentrator (CEC) \cite{Winston:2005}. Simulations using the ZEPHYROS code \cite{Robinson:2012} will provide insight into electron transport and heating inside these targets. The predictions of the theory presented in section \ref{SecEtendue} will be compared with the results of the simulation.

Section \ref{SecResults} will summarise the results and present the conclusions.

\section{Fast electron etendue \label{SecEtendue}}

In this section we will provide a short introduction of the concept of etendue and maximum concentration in optics. The specular reflection of fast electrons at material interfaces and the ballistic motion inside the homogeneous domains allows us to treat the fast electron trajectories the same way as light rays in geometric optics. Specifically we can apply the same fundamental conservation laws to the electron dynamics. The etendue of a source which is extended over a small surface area $dS$ with rays confined to a small solid angle $d\Omega$ is given by
\begin{equation}
dG = n^2dS\,cos\vartheta\,d\Omega,
\end{equation}
where $\vartheta$ is the angle that the rays make with the normal to the surface element and $n$ is the refractive index of the medium. When translating the etendue concept to fast electron motion, the notion of a refractive index loses its meaning. The deflection of an electron as it passes through a magnetic field layer does not obey Snell's law. We will therefore be limiting ourselves to a single material and the specular reflections of electrons at the magnetic field layers. This implies that we can set the refractive index to $n=1$ and omit it from the calculation.

Given a finite sized circular source $S$ with radius $R$ emitting rays with an angular spread of opening angle $\theta$, the total etendue of the source is
\begin{equation}
G = \int_0^{2\pi}\int_0^R\int_{\Omega_{\theta}}r\cos\theta\;d\Omega\,dr\,d\phi = \pi^2 R^2 \sin^2\theta.
\end{equation}
Here $\Omega_{\theta}$ is the solid angle of the cone with half angle $\theta$. One can show directly from Liouville's theorem that the etendue of a bundle of rays is a conserved quantity. This means that, given a second circular surface $S'$ with radius $R'$, and assuming that all rays from the source pass through $S'$ then $G'=G$, or
\begin{equation}
R' \sin\theta' = R \sin\theta.
\end{equation}
Here $\theta'$ is the cone half angle of the rays passing through $S'$. The maximum value that the angle of a bundle of forward going rays can assume is $\theta_{\text{max}} = \pi/2$. This leads to a minimum value of $R'$ or a maximum concentration,
\begin{equation}
R'_{\text{min}} = R \sin\theta. \label{EqRmin}
\end{equation}
The product $R\sin\theta$ is a measure of how well a bundle of rays can be collected and directly gives the minimum radius of a disk that this bundle can be focused onto. 

In general the angular spread of fast electrons will depend on many parameters, such as the laser intensity or the laser contrast ratio. Green et al \cite{Green:2008} compared experimental data with other results published in the literature and concluded that the angular spread increases with laser intensity. This means that, assuming a fixed power of the laser beam, the angular spread of fast electrons increases as the beam radius decreases. Thus, the beam radius and the angular spread are competing factors in an attempt to minimise $R'_{\text{min}}$ and it is not a priori clear whether a small spot size coupled with a large angular spread is better than a large spot size coupled with a small angular spread. In order to make a prediction we use a crude fit to the data presented by Green et al \cite{Green:2008},
\begin{equation}
\theta = 17.3\degree + 5.85\degree \ln\frac{I}{10^{18}\text{W cm}^{-2}}. \label{EqThetaApprox}
\end{equation}
This formula is based on intensities in the range from $10^{18}\text{W cm}^{-2}$ to $10^{21}\text{W cm}^{-2}$.  It should be emphasised that the form above is only a rough approximation to a small data set. We use it here merely to derive an approximate functional dependence of $R'_{\text{min}}$ on the spot radius $R$. For fixed beam power $P$ the average intensity over a circular disk is
\begin{equation}
I = \frac{P}{\pi R^2}. \label{EqPower}
\end{equation}
This relation allows us to express $\theta$, and therefore $R'_{\text{min}}$, purely as a function of the beam radius.

\begin{figure}
\begin{center}
\includegraphics[width=10cm]{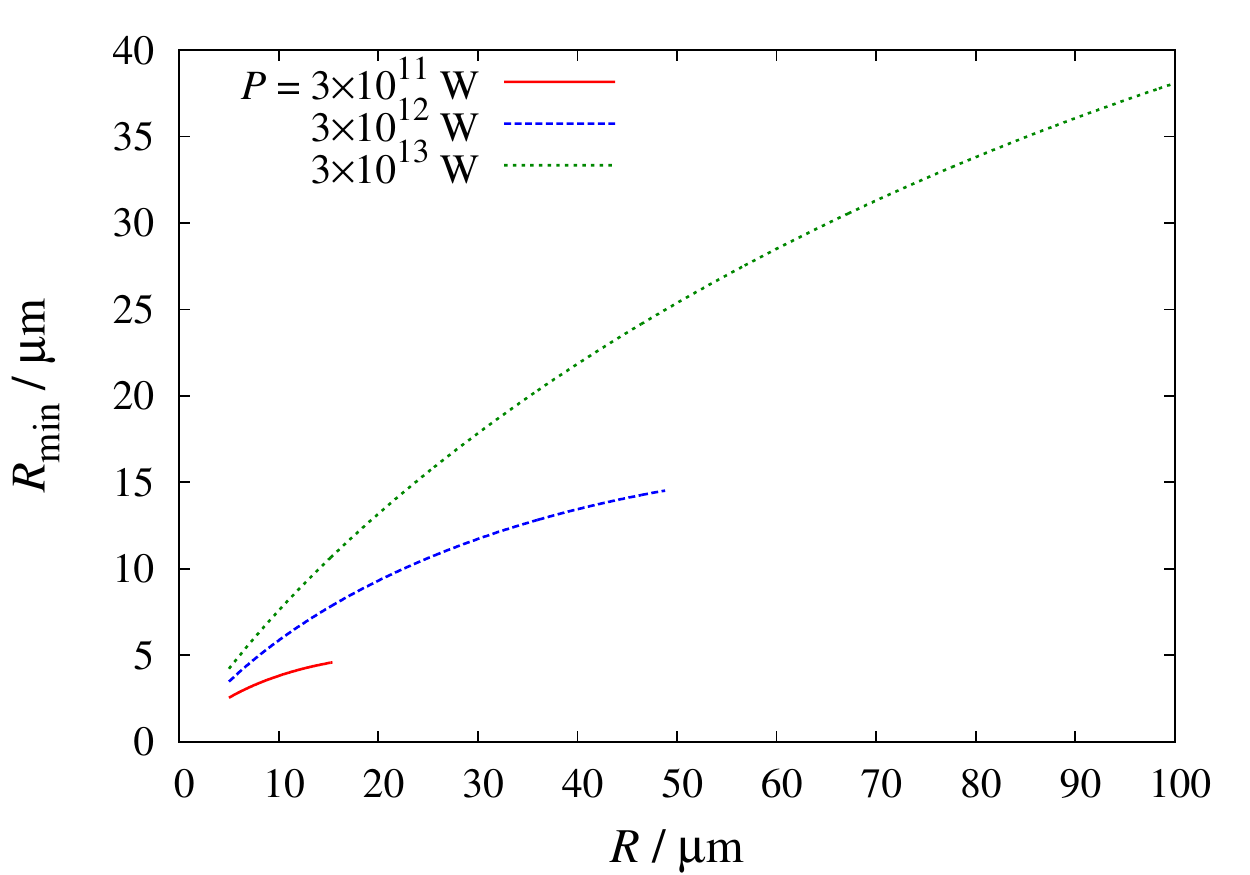}
\end{center}
\caption{Minimum radius of fast electron beam as a function of laser spot size for various laser beam powers. The curves are limited to the region of validity of equation \ref{EqThetaApprox}, between $I = 10^{18}\text{W cm}^{-2}$ and $10^{21}\text{W cm}^{-2}$.}
\label{FigRminVsR}
\end{figure}

Figure \ref{FigRminVsR} shows the dependence of $R'_{\text{min}}$ on the beam Radius $R$ for various powers. One can see that, due to the logarithmic dependence of $\theta$ on the intensity, the advantage of decreasing the spot radius outweighs the disadvantage of an increasing angular spread of fast electrons. The minimum achievable focusing radius increases with the power of the incident laser, as is to be expected. In addition, $R'_{\text{min}}$ also increases with $R$ despite the fact that that angular spread decreases. For smaller values of $P$ the curves plotted in figure \ref{FigRminVsR} stop at some maximum value of $R$. This corresponds to the radius at which the intensity drops below $I = 10^{18}\text{W cm}^{-2}$, below which equation \ref{EqThetaApprox} loses its justification. Although the curves seem to flatten towards this point, we expect the value of $R'_{\text{min}}$ to steadily increase with $R$. This is due to the fact that for low intensities the value of $\theta$ will not fall below a finite positive lower bound. Thus, the value of $\sin\theta$ is also bounded from below implying that $R\sin\theta$ will grow with $R$.

The results of this section seem to imply that it is always favourable to focus the laser spot to the smallest radius possible. This will result in a point linke source of fast electrons with a large angular spread. Given that the $\theta$ is less than $90\degree$ the value of $R'_{\text{min}}$ will be below $R$ which means that the electrons can be guided into a region even smaller than the laser spot size where they can rapidly heat the target.

It should be pointed out that the values of $R'_{\text{min}}$ presented here impose a theoretical lower limit to the size of the heated structure. Whether this limit can be achieved in a concrete experiment design will depend on other factors. As we will see in the next section, the losses of fast electrons through the magnetic field layer play an important role in designing the optimal target geometry. These losses are not included in the calculations above in which we assumed perfectly reflecting interfaces.

\section{Simulations \label{SecSimulations}}

\begin{figure}
\begin{center}
\includegraphics[width=7cm]{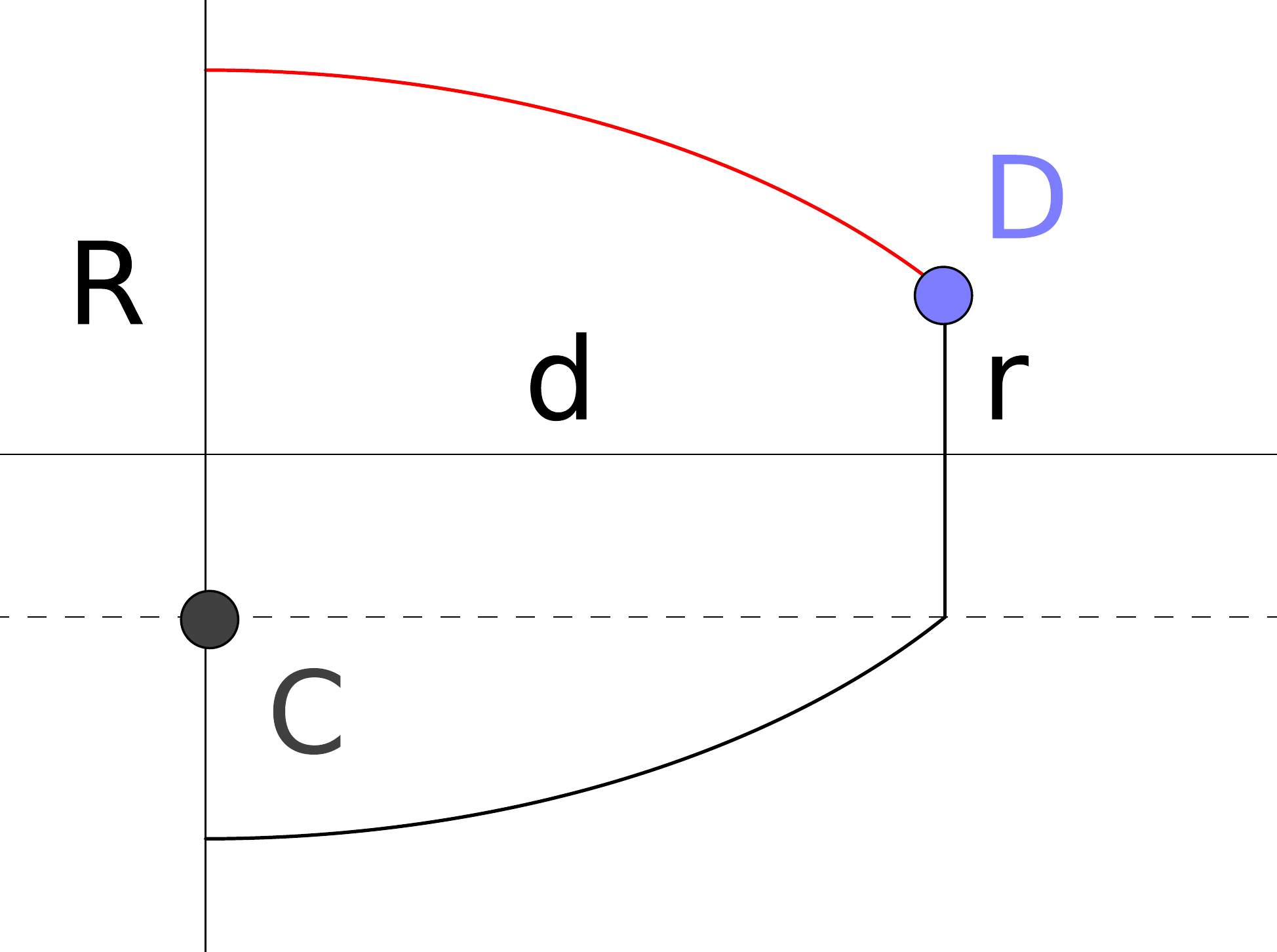}
\end{center}
\caption{Geometry of the target used in the simulations.}
\label{FigConcentrator}
\end{figure}

In order to evaluate the theoretical calculations we performed simulations of the electron transport in structured targets. In contrast to imaging optics, a non-imaging component does not need to map each point on some object plane uniquely onto a point on an image plane. Instead, different rays from a single point on the source may end up on different points on a receiver, depending on their path. While designing an ideal concentrator using imaging optics is highly impractical, using non-imaging optics the task is greatly simplified. Various methods are used to design non-imaging components \cite{Winston:2005}. In the {\em Edge Ray Approach} limiting rays originating from the edges of the source are considered and the design is chosen so that these rays hit the edges of the receiver. Geometric arguments are used to prove that rays emerging from between the edges of the source will end up between the edges of the receiver. The {\em Flow Line Approach}, on the other hand, is more general and uses the description of the Hamiltonian flow in 6 dimensional space.

The most common designs resulting from these methods are the Compound Parabolic Concentrator (CPC) and the Compound Elliptical Concentrator (CEC). In these designs a section of a parabola or ellipse is rotated around the optical axis to create a reflecting surface. Here the symmetry axis of the parabola or ellipse is not identical to the axis of rotation and, in general, does not have to be aligned with it. While CPCs are used for infinitely distant sources (parallel rays), CECs are used for finite sized diffuse sources. 

In our simulations we used a variant of a CEC. The geometry of the collector is shown in figure \ref{FigConcentrator}. The simulations were carried out using the 3D particle hybrid code ZEPHYROS \cite{Robinson:2012}. In all simulations presented here the bulk of the target consists of CH while the plug that acts as the collector is made from carbon. The radius of the plug at the target front surface is $R$ and the small radius inside the target is $r$ where the end surface is located. The sides of the plug are defined by a section of an ellipse that is shifted down along the $y$--axis by the small radius $r$, denoted by the red curve in figure \ref{FigConcentrator},
\begin{equation}
\frac{x^2}{a^2} + \frac{(y+r)^2}{b^2} = 1 \quad \text{with} \quad y>0, 0\le x \le d,
\end{equation}
where $d$ is the depth of the plug and $a$ and $b$ are the semi-major and the semi-minor axes respectively. The centre $C$ of the ellipse lies on the front surface and the semi-minor axis has a radius of $b=R+r$. The semi-major axis is chosen such that the ellipse passes through the point $D = (d,r)$,
\begin{equation}
a = \frac{bd}{\sqrt{b^2-4r^2}}.
\end{equation}
This section is then rotated around the $x$--axis to create the rotationally symmetric collector. Note that typical CEC geometries prescribe the locations of the focal points in order to optimise the number of rays collected onto the end surface. Here we do not impose such a restriction due to a number of other constraints imposed by the nature of the fast electron transport. Firstly we don't expect the fast electrons to travel arbitrary distances inside the solid density target before losing their energy. This means that we want to retain the depth $d$ as one of the free parameters. Fixing the focal points implies that $d$ is a function of $r$ and $R$ and can not be chosen freely. Secondly our geometry ensures that the ellipse intersects the front surface of the target at a right angle. This minimises the losses of fast electrons through the material interface.

\begin{table}[]
\begin{center}
\renewcommand{\arraystretch}{1.2}
\begin{tabular}{|l|l|l|l|l|l|l|l|}
\hline
Run & \makecell{Power\\ \hline $10^{13}$W} & \makecell{$R_{\text{spot}}$ \\ \hline $\mu\text{m}$} & \makecell{$I/10^{18}$ \\ \hline $\text{W\;cm}^{-2}$} & \makecell{$R$ \\ \hline $\mu\text{m}$} & \makecell{$r = R_{\text{min}}$ \\ \hline $\mu\text{m}$} & $\theta$ & \makecell{$d$ \\ \hline $\mu\text{m}$}\\ \hline
A$_{30}$, A$_{40}$, A$_{50}$   & 8                  & 5                               & 101.86                      & 10               & 6.99                              & 0.774   & 30, 40, 50 \\ \hline
B$_{30}$, B$_{40}$, B$_{50}$   & 8                  & 10                              & 25.47                       & 20               & 11.823                            & 0.632   & 30, 40, 50 \\ \hline
C$_{30}$, C$_{40}$, C$_{50}$   & 8                  & 20                              & 6.37                        & 40               & 18.858                            & 0.491   & 30, 40, 50 \\ \hline
D$_{30}$, D$_{40}$, D$_{50}$   & 2                  & 5                               & 25.47                       & 10               & 5.9114                            & 0.632   & 30, 40, 50 \\ \hline
E$_{30}$, E$_{40}$, E$_{50}$   & 2                  & 10                              & 6.37                        & 20               & 9.4290                            & 0.491   & 30, 40, 50 \\ \hline
F$_{30}$, F$_{40}$, F$_{50}$   & 2                  & 20                              & 1.5915                      & 40               & 13.693                            & 0.349   & 30, 40, 50 \\ \hline
\end{tabular}
\end{center}
\caption{Overview of the parameters of the ZEPHYROS simulation runs.}
\label{tblSimulationParams}
\end{table}

We performed two sets of runs, each with fixed power but varying laser spot sizes. The first set of runs (A$_d$, B$_d$, and C$_d$) used a laser power of $P=8\times 10^{13}$W, the second set (runs D$_d$, E$_d$, and F$_d$) used $P=2\times 10^{13}$W. For each power three laser spot sizes, $R_{\text{spot}}=5$, $10$, and $20\mu$m were considered. For each run the average intensity  and the fast electron spread were calculated using equations \ref{EqPower} and \ref{EqThetaApprox}. In all cases the front radius of the collector was chosen to be twice the laser spot radius in order to increase the number of fast electrons captured. Using equation \ref{EqRmin} with the front collector radius, the minimum radius $R_{\text{min}}$ was determined. This value was used for the small radius of the collector $r = R_{\text{min}}$. In each case, three runs with depth $d = 30, 40, $ and $50\mu$m were performed. An overview of the parameters can be seen in table \ref{tblSimulationParams}.

In all simulation runs a box size of $300\times300\times200$ was chosen. The cell size was $\Delta x=0.5\mu$m, resulting in a box of size $150\times150\times100\mu\text{m}^2$. The laser was incident on the left boundary for 1ps and the total duration of the simulation was 2ps. The beam has a Gaussian transverse profile with a radius of $R_{\text{spot}}$ and an intensity of $I$ as given in table \ref{tblSimulationParams}. The laser is not simulated directly but is used to inject fast electrons into the simulation domain. The electrons have an exponential energy distribution with a temperature of $T_f$ given by Wilks' scaling \cite{Wilks:1992}
\begin{equation}
T_f = 2m_ec^2\left( \sqrt{1+\frac{I\lambda_L^2}{1.38\times10^{18}}} - 1\right),\label{EqWilksScaling}
\end{equation}
where $I$ is taken in units of $10^{18}\text{W\;cm}^{-2}$ and the laser wavelength $\lambda_L$ is given in units of $\mu\text{m}$. The resistivity of the target is calculated using the model of Lee and More \cite{Lee:1984}.

\section{Results\label{SecResults}}

\begin{figure}[ht]
\begin{center}
\includegraphics[width=0.33\textwidth]{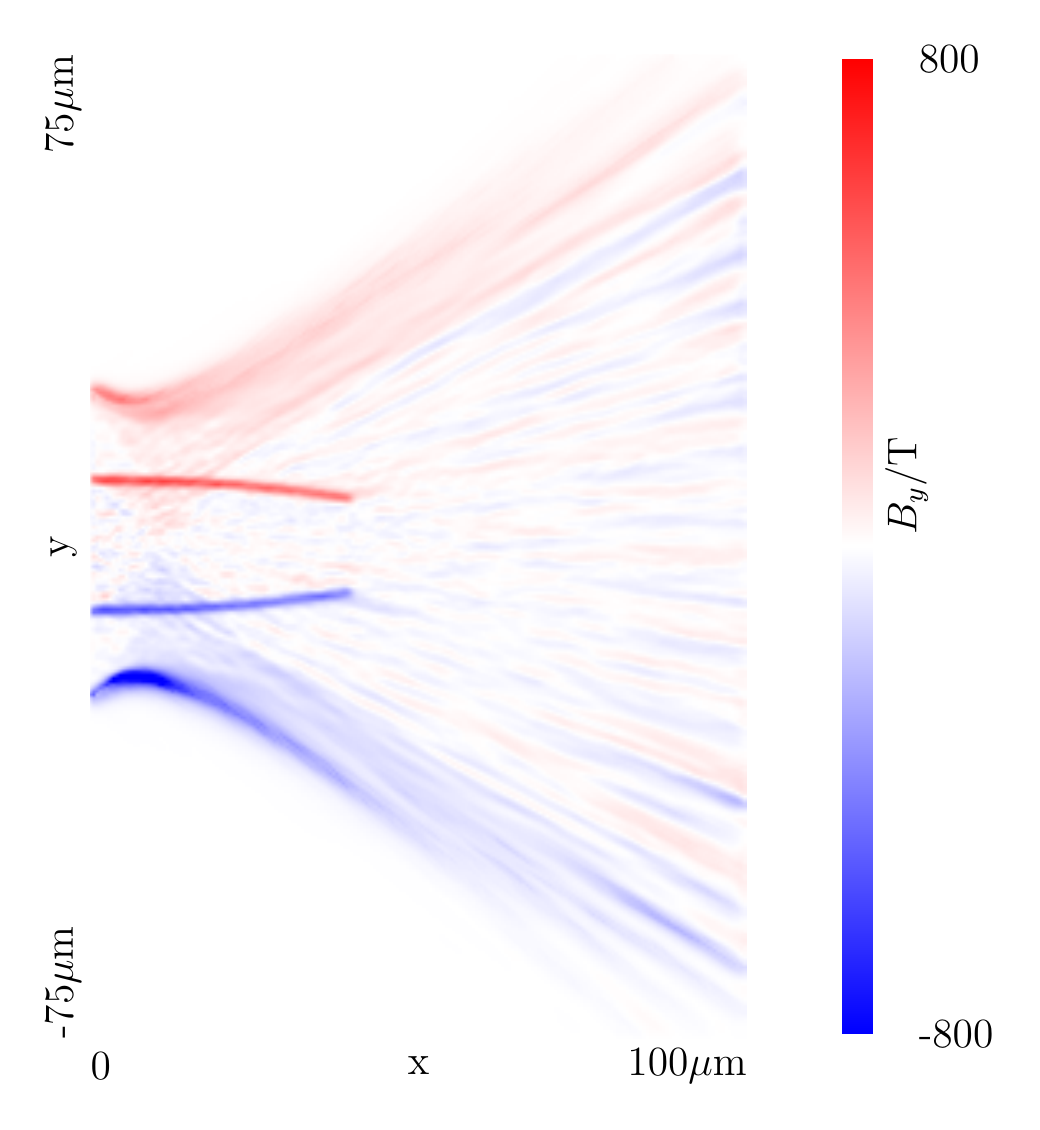}\hfill
\includegraphics[width=0.33\textwidth]{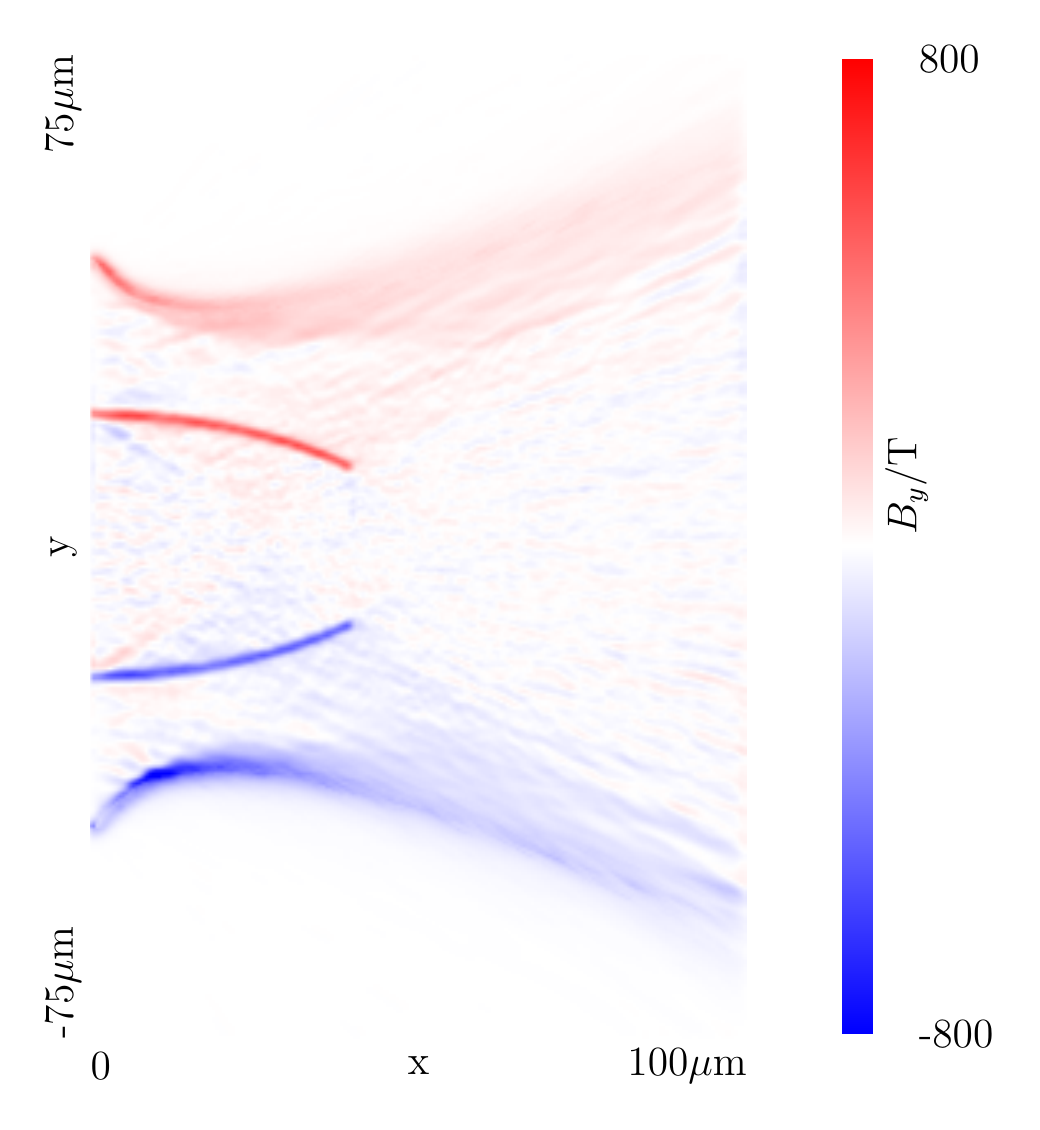}\hfill
\includegraphics[width=0.33\textwidth]{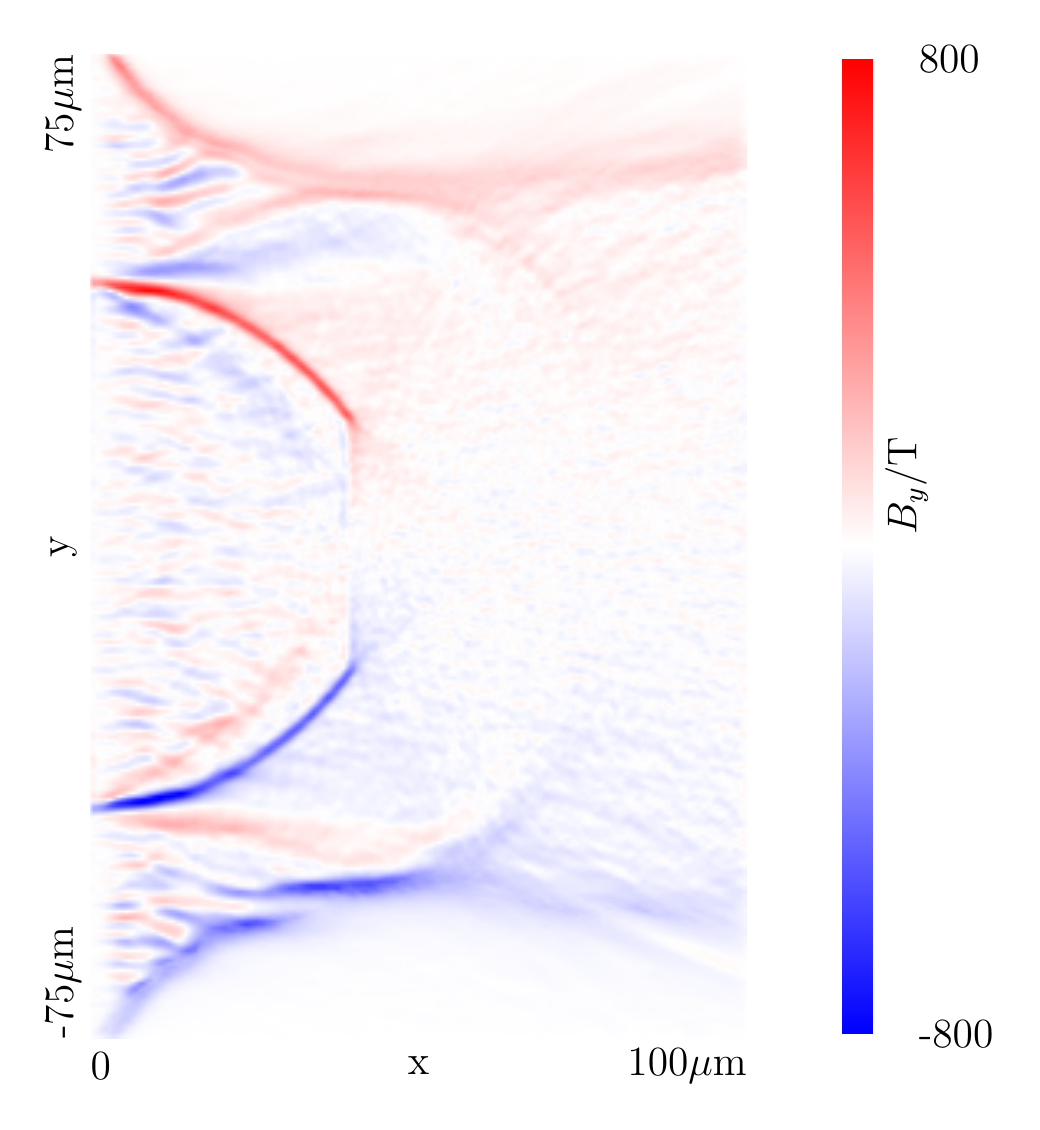}
\end{center}
\caption{The out of plane magnetic field $B_y$ at the end of the simulation. The length of the collector is $40\mu$m and the power is $P=8\times10^{13} W$. The laser spot radius is $R=5\mu$m (run A$_{40}$, left panel), $10\mu$m (run B$_{40}$, middle panel), and $20\mu$m (run C$_{40}$, right panel). The opening radius of the collector is twice the laser spot radius. The electron divergence angle is calculated from equation \ref{EqThetaApprox} and the small radius of the collector is chosen to be equal to $R'_{\text{min}}$ given by equation \ref{EqRmin}}
\label{FigBySurface}
\end{figure}

Figure \ref{FigBySurface} shows the out of plane magnetic field $B_y$ for the runs A$_{40}$, B$_{40}$, and C$_{40}$ with $R_{\text{spot}}=5$, $10$, and $20\mu$m respectively at the end of the simulation. In all cases the sharp resistivity gradient at the material interface results in magnetic field strengths of around 600 to 800T. Note that no magnetic field is observed at the circular tip of the concentrator. This is to be expected since the fast electron current here is perpendicular to the surface, and only the parallel component of the fast electron current generates a magnetic field. At larger radii another, more diffuse magnetic field is observed. This field is caused by the radial gradient of the fast electron current. The radial location of the field scales with the radius of the laser spot size. Due to the angular spread of the electrons this field is not able to contain the fast electrons.

\begin{figure}[ht]
\begin{center}
\includegraphics[width=0.33\textwidth]{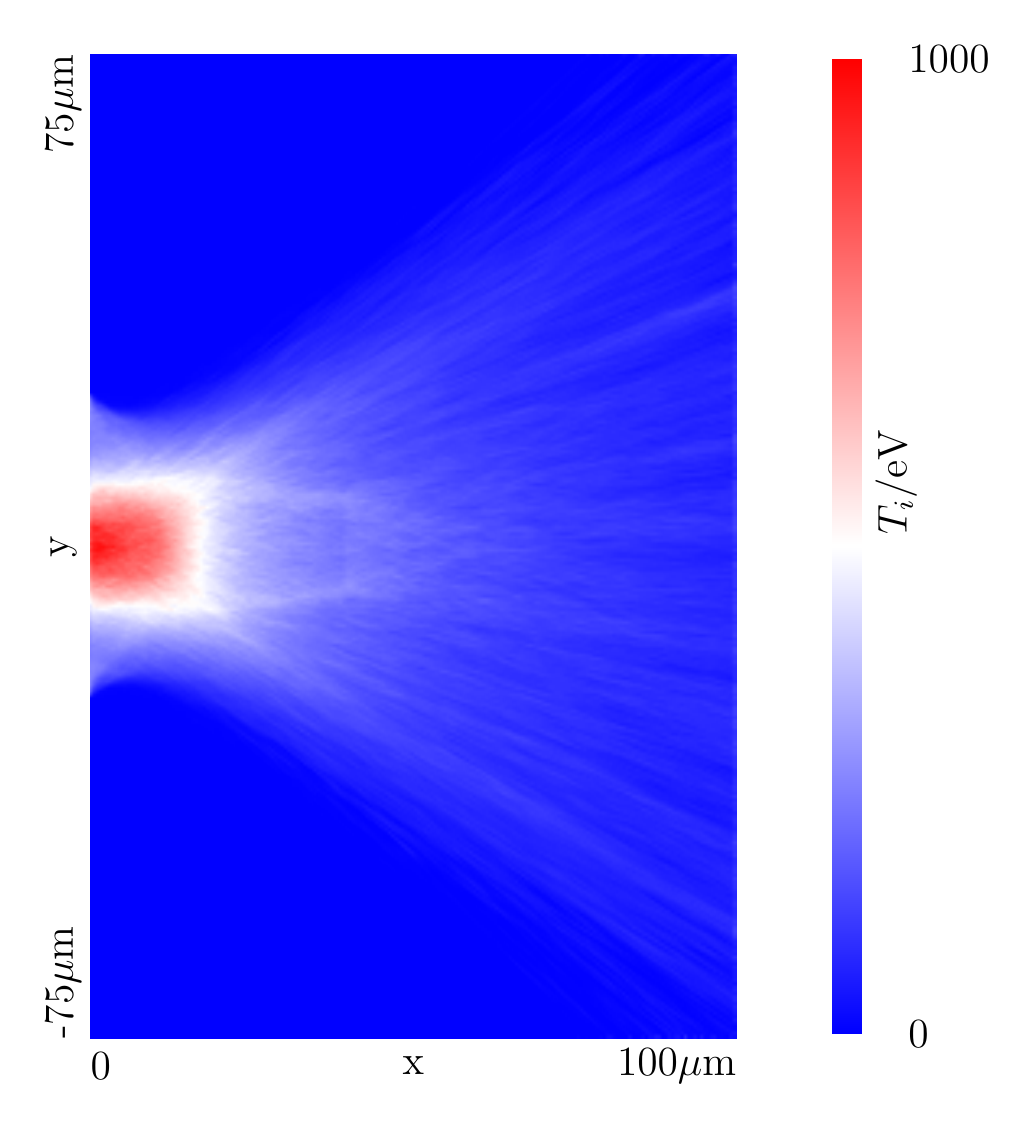}\hfill
\includegraphics[width=0.33\textwidth]{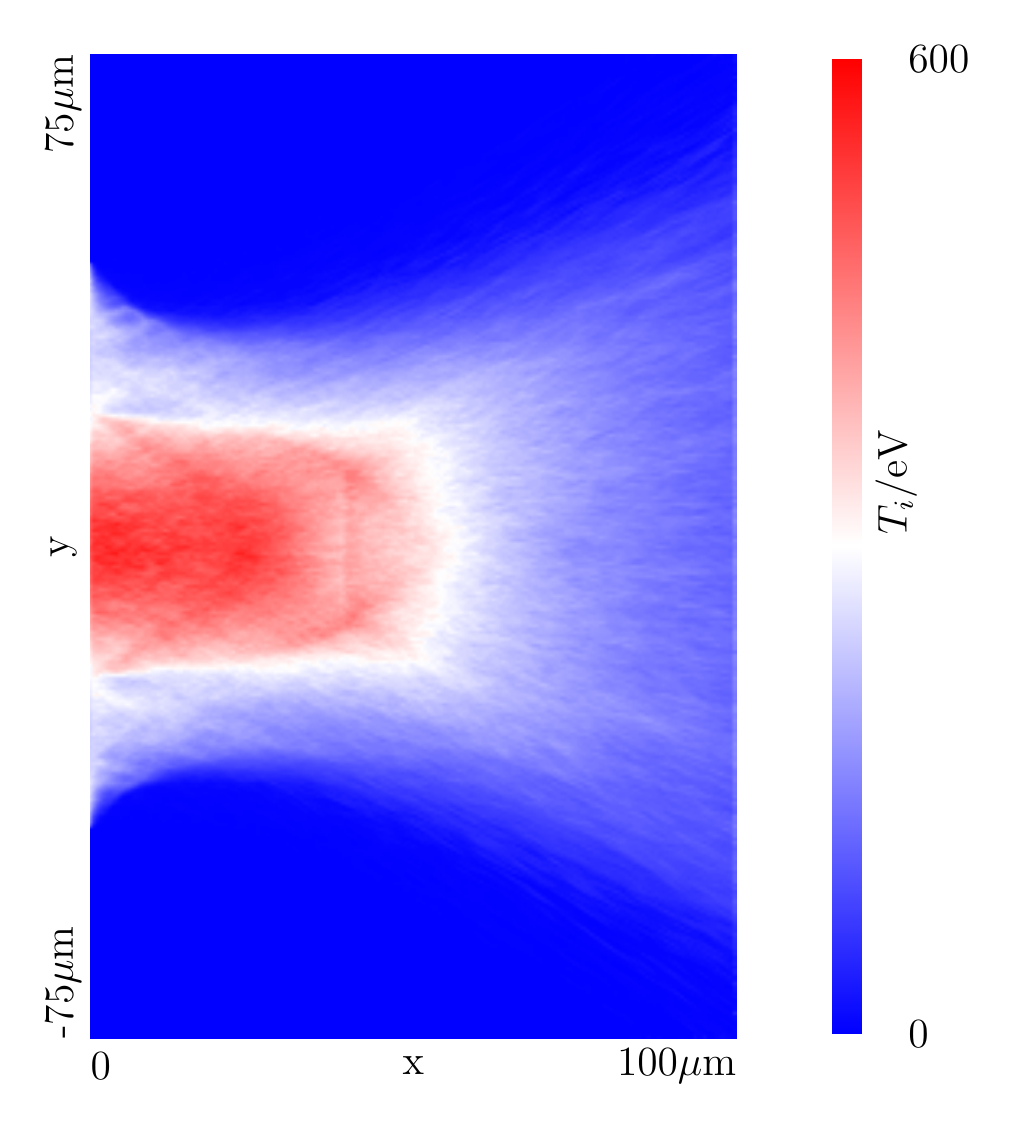}\hfill
\includegraphics[width=0.33\textwidth]{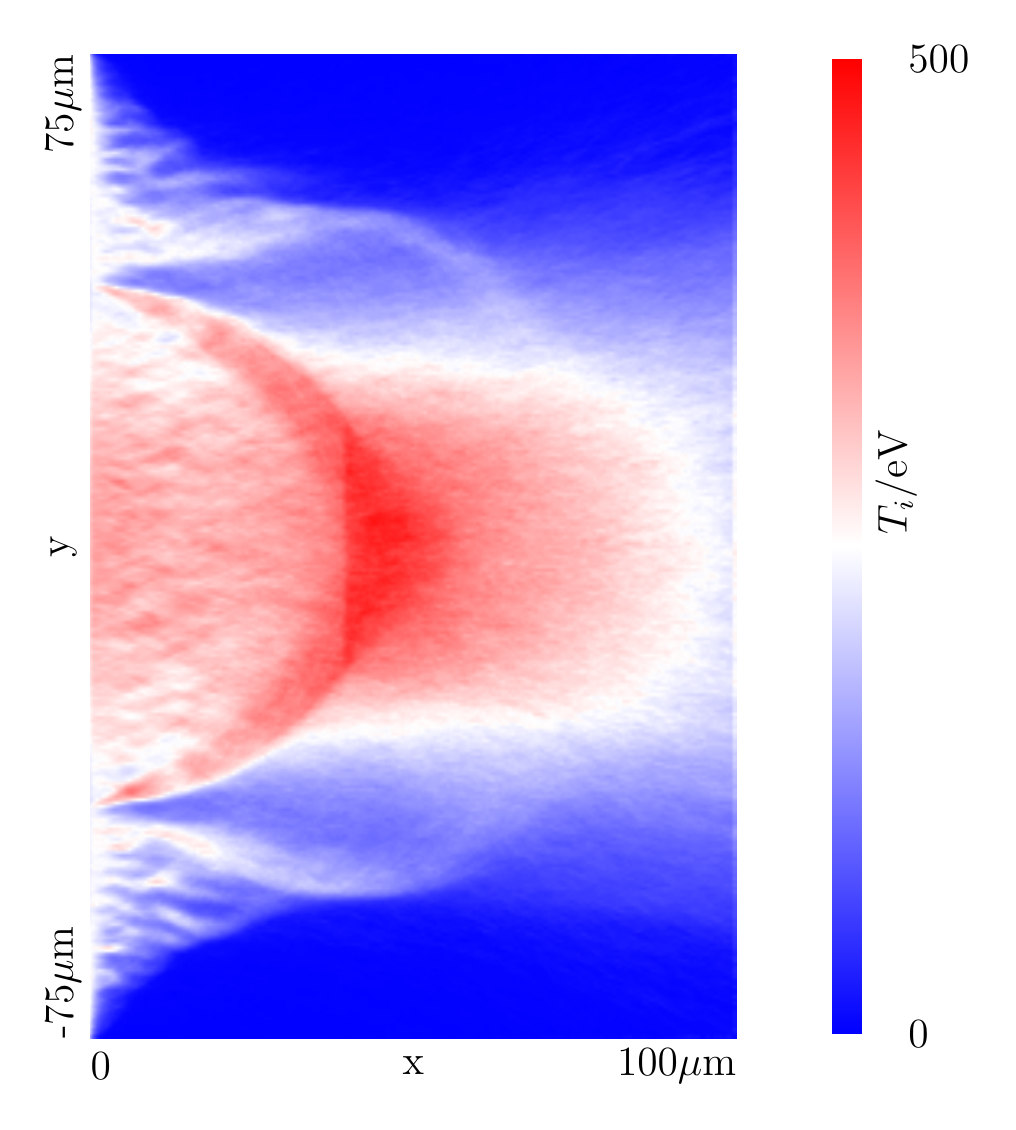}
\end{center}
\caption{The ion temperature at the end of the simulation for runs A$_{40}$ (left panel), B$_{40}$ (middle panel), and C$_{40}$ (right panel).}
\label{FigTiSurface}
\end{figure}

\begin{figure}[ht]
\begin{center}
\includegraphics[width=0.33\textwidth]{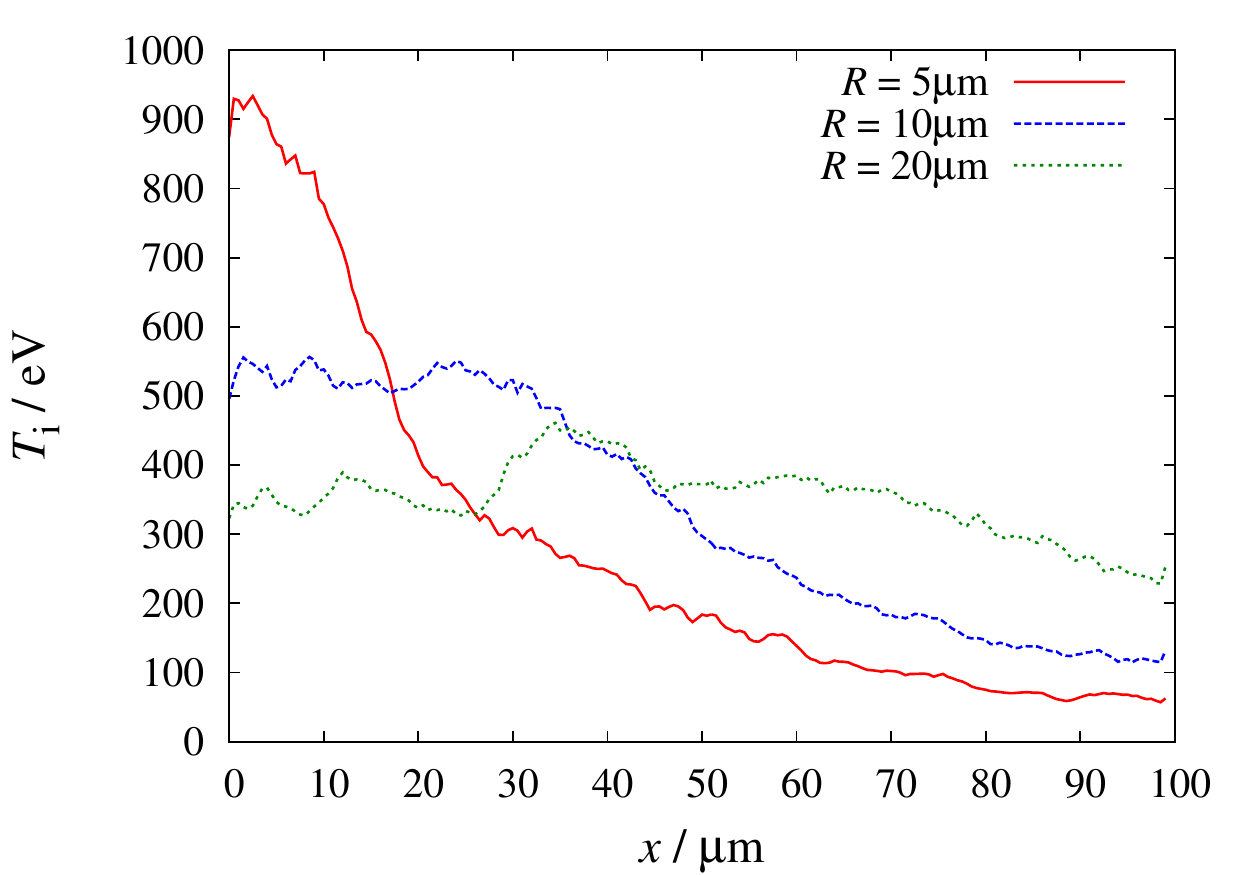}\hfill
\includegraphics[width=0.33\textwidth]{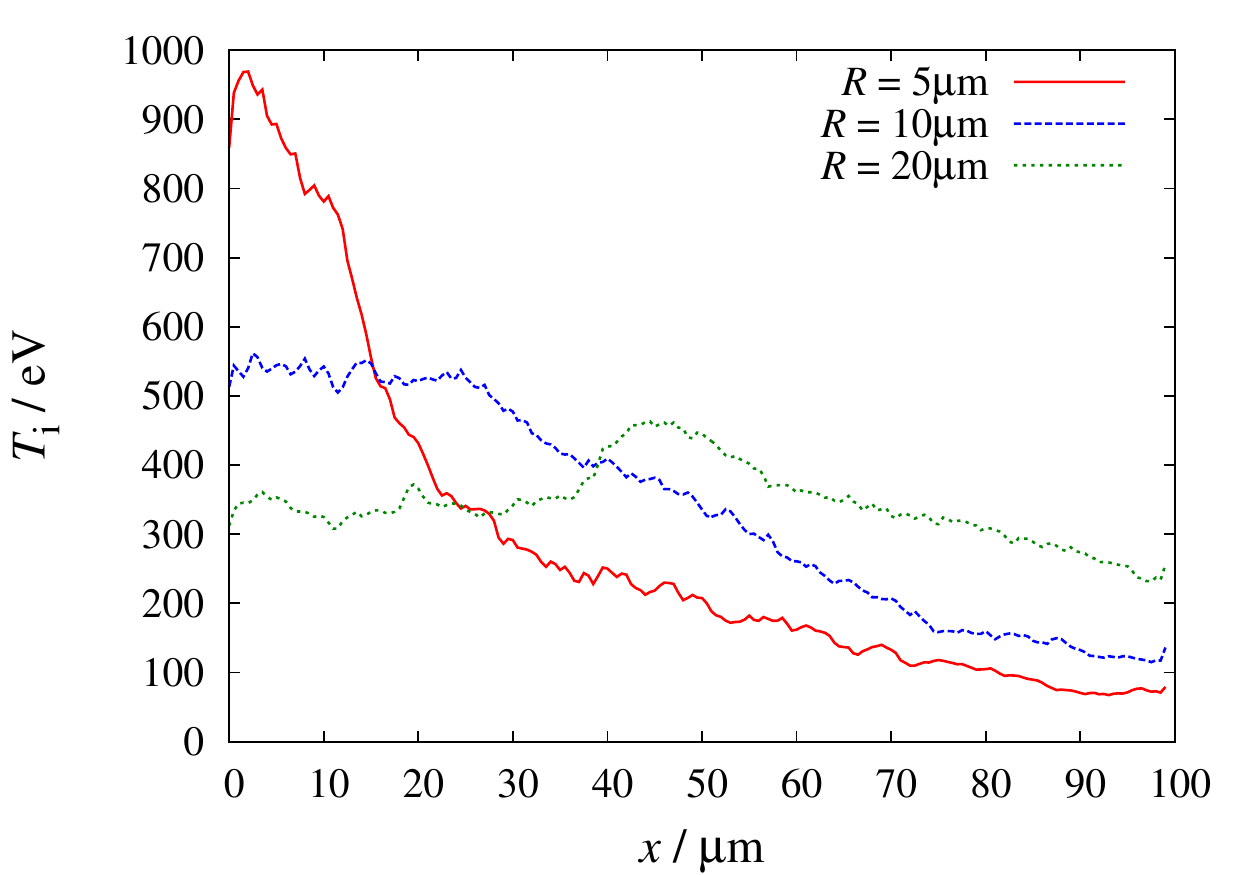}\hfill
\includegraphics[width=0.33\textwidth]{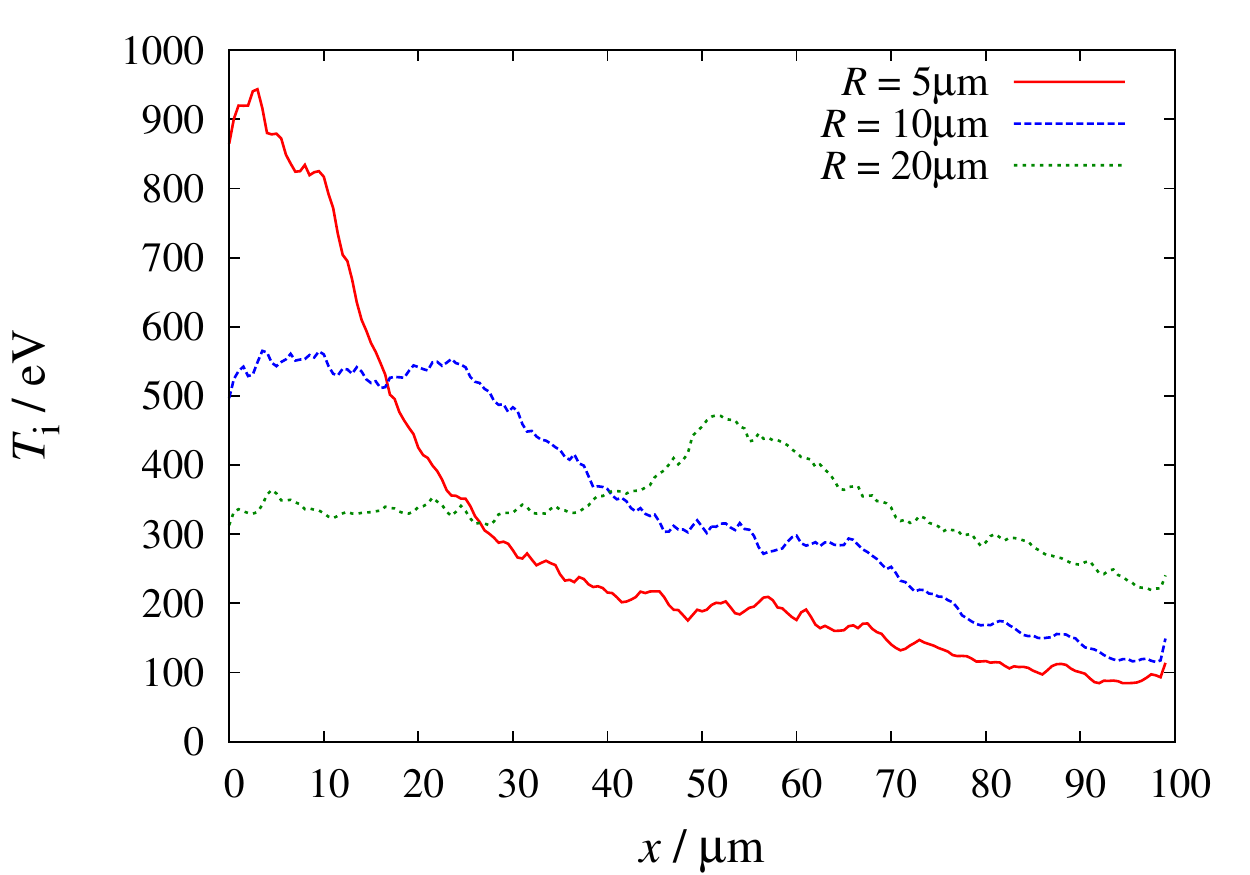}
\end{center}
\caption{The ion temperature at the end of the simulation along the simulation axis for different spot radii $R$. The power is $P=8\times10^{13} W$. The length of the collector is $L=30\mu$m (left panel), $40\mu$m (middle panel), and $50\mu$m (right panel). The opening radius of the collector is twice the laser spot radius. The electron divergence angle is calculated from equation \ref{EqThetaApprox} and the small radius of the collector is chosen to be equal to $R'_{\text{min}}$ given by equation \ref{EqRmin}}
\label{FigTiAxis}
\end{figure}

Figure \ref{FigTiSurface} shows the ion temperature $T_i$ inside the target for a slice $z=0$ for runs A$_{40}$, B$_{40}$, and C$_{40}$ . In section \ref{SecEtendue} it was shown that a smaller spot radius implies a smaller etendue value and hence better focusability onto a small surface. One might anticipate that smaller etendue will improve the ability to deposit the electron energy into a small volume inside the target. Comparing the three panels of figure \ref{FigTiSurface} one can see that the overall heated volume indeed increases with spot radius. However, for the smallest radius of $R_{\text{spot}}=5\mu$m (run A$_{40}$) the energy deposited is concentrated in the first 15$\mu$m near the front surface of the target, at $x=0$. There is almost no heating present near the tip of the collector geometry. Traces of fast electrons escaping the carbon plug at large angles can be seen beyond 15$\mu$m inside the target. This indicates that fast electrons are escaping the concentrator. Due to the high intensity of the laser beam for this power and at this spot radius the angular spread has a half angle of around $44\degree$. In addition, higher intensities correspond to higher electron energies according to equation \ref{EqWilksScaling}. Both factors reduce the ability of the self generated magnetic field at the carbon--CH material boundary to reflect the electrons. In figure \ref{FigTiAxis} the ion temperature $T_i$ is plotted along the $x$--axis for the runs A$_d$, B$_d$ and C$_d$ with $d=30\mu$m (left panel), $d=40\mu$m (middle panel), and $d=50\mu$m (right panel). For all three values of $d$ the temperature of runs A$_d$ show similar profiles with a maximum of over 900eV near the front surface and quickly falling off to below 300eV for $x\gtrsim 30\mu$m.

For a laser spot radius of $R_{\text{spot}}=10\mu$m, run B$_{40}$, the number of escaping electrons is greatly reduced and the carbon plug is heated more evenly and to greater depths, as can be seen in the middle panel of figure \ref{FigTiSurface}. With a maximum around 550eV the ion temperature is lower than what was seen for run A$_{40}$ . However, the heated region extends all the way to the tip of the collector and beyond. Two feature are visible. Inside the collector region a heated structure converging from the edge of the collector at the front surface towards the centre near $x\approx25\mu$m. This feature seems to be caused by fast electrons being redirected by the magnetic field just as they are injected at the front surface. A second feature can be seen at the tip of the collector and beyond. An increased temperature is observed at the material interface around the outer edge for $x>30\mu$m and in the direct extension of this interface up to $x\approx 50\mu$m in the CH cladding. The heating in this region is caused by fast electrons being guided along the outer edge of the collector and subsequently leaving the tip at a tangential direction. The on-axis temperature profiles for runs B$_d$, $d=30, 40, 50\mu$m in figure \ref{FigTiAxis} show little dependence on the length of the collector. In all three cases the temperature profile is almost flat around 530eV up to a depth of $x\approx 30\mu$m. Deeper into the target, the temperature falls off. The rate of decrease is smaller for the $R_{\text{spot}}=10\mu$m runs compared with the $R_{\text{spot}}=5\mu$m runs and for $x\gtrsim20\mu$m the ion temperature for runs B$_d$ is larger than the ion temperature for runs A$_d$, independent of the value of $d$.

When the laser spot is increased to $R_{\text{spot}}=20\mu$m in run C$_{40}$ the two features that are observed in run B$_{40}$ are still present but they now overlap due to the modified aspect ratio of the collector geometry, as can be seen in the right panel of figure \ref{FigTiSurface}. The heated region originating from the edge of the collector at the front surface converges to a point just beyond the tip of the collector around $x\approx45\mu$m. At the same time electrons are guided tangentially along the outer edges of the geometry and leave the collector to converge in the same region. This results in an increase in heating just beyond the tip, inside the CH material. The temperature profiles along the $x$--axis shown in figure \ref{FigTiAxis} confirm this behaviour. For all three values of $d$ the temperature profile is flat around 350eV for small values of $x$. An increase in temperature can be observed inside the target and the location of the maximum depends on the value of $d$. This indicates that the guiding structure of the collector is indeed responsible for the increase in heating deep inside the target.

\begin{figure}[t]
\begin{center}
\includegraphics[width=0.33\textwidth]{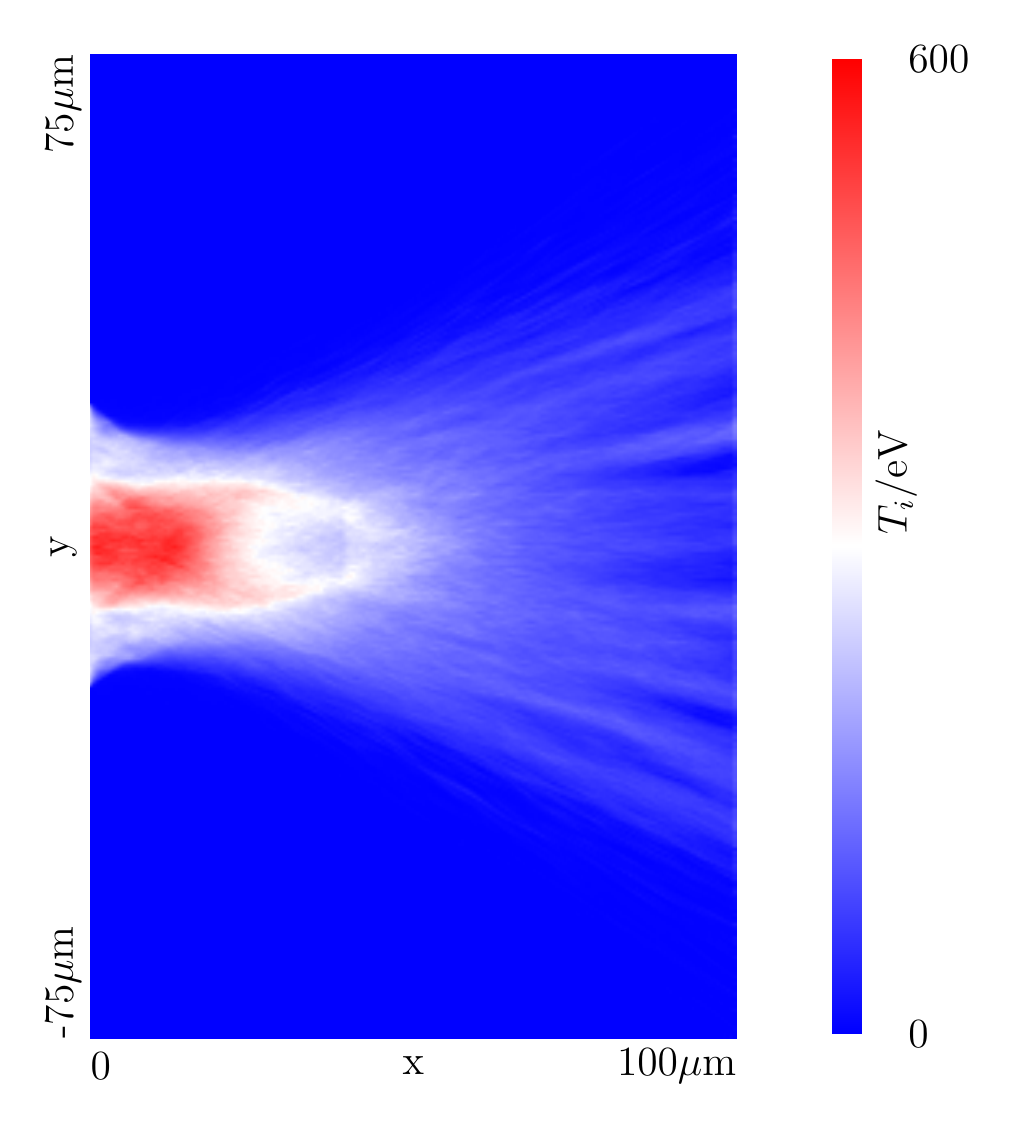}\hfill
\includegraphics[width=0.33\textwidth]{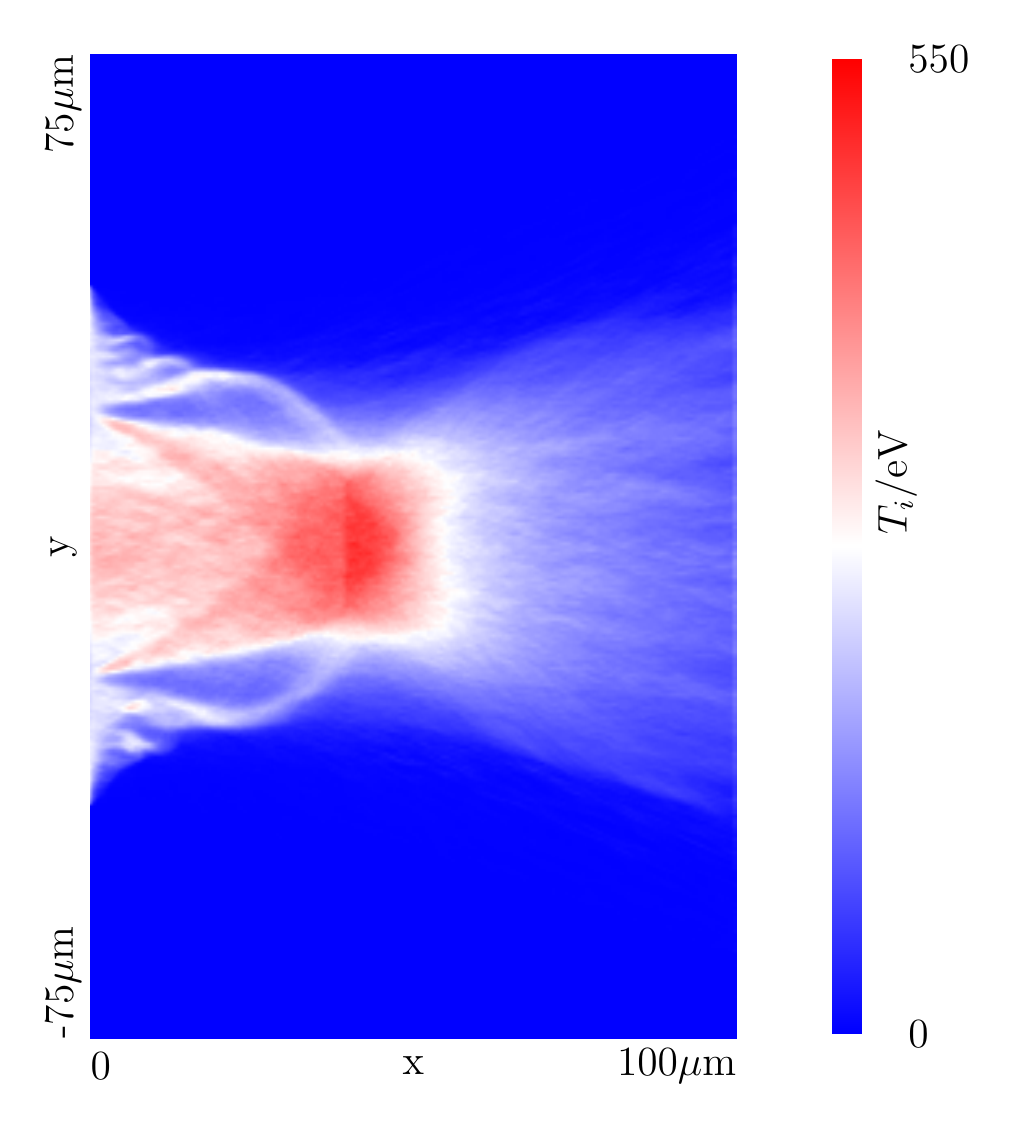}\hfill
\includegraphics[width=0.33\textwidth]{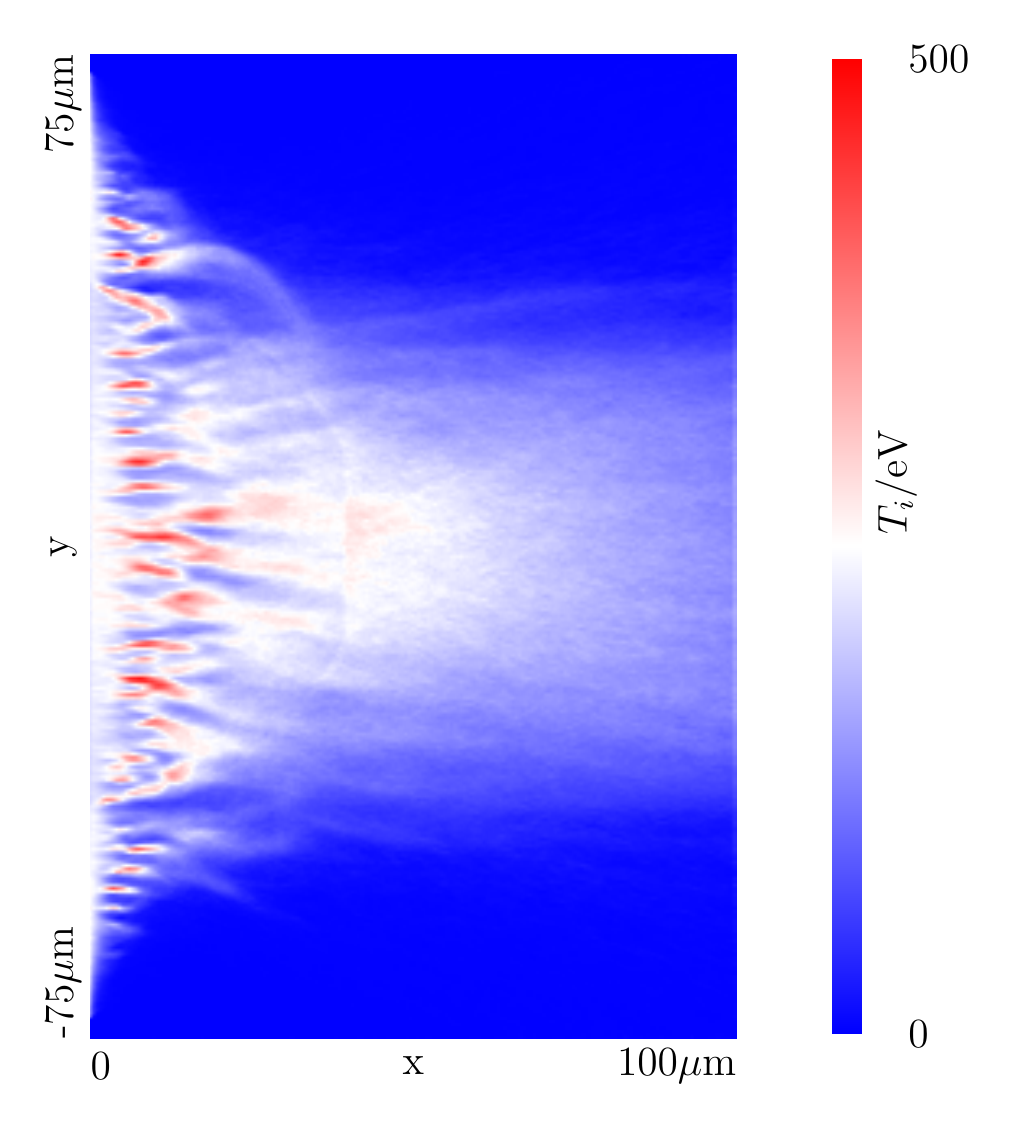}
\end{center}
\caption{The ion temperature at the end of the simulation. The length of the collector is $40\mu$m and the power is $P=2\times10^{13} W$. The laser spot radius is $R=5\mu$m (run D$_{40}$, left panel), $10\mu$m (run E$_{40}$, middle panel), and $20\mu$m (run F$_{40}$, right panel). The opening radius of the collector is twice the laser spot radius. The electron divergence angle is calculated from equation \ref{EqThetaApprox} and the small radius of the collector is chosen to be equal to $R'_{\text{min}}$ given by equation \ref{EqRmin}}
\label{FigTiSurfaceLowPower}
\end{figure}

\begin{figure}[ht]
\begin{center}
\includegraphics[width=0.33\textwidth]{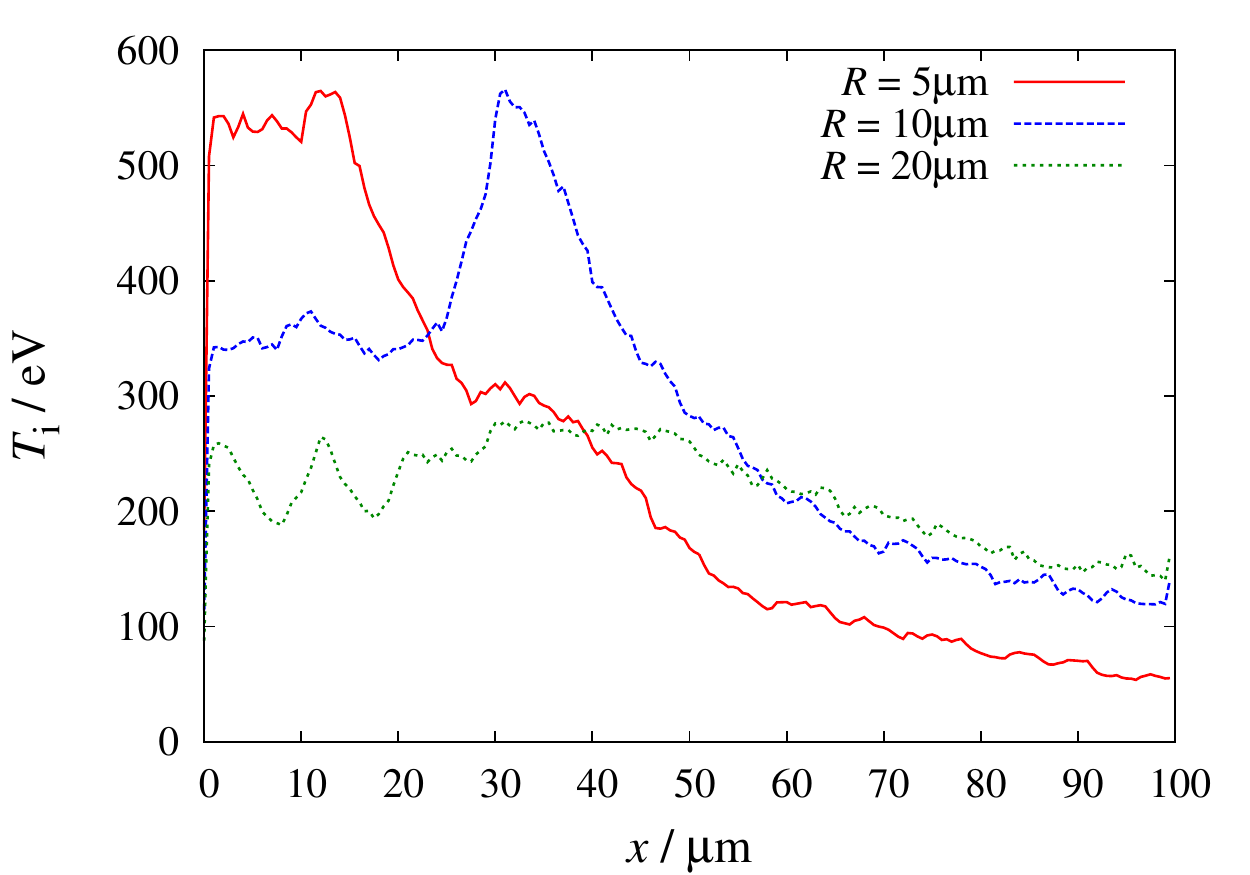}\hfill
\includegraphics[width=0.33\textwidth]{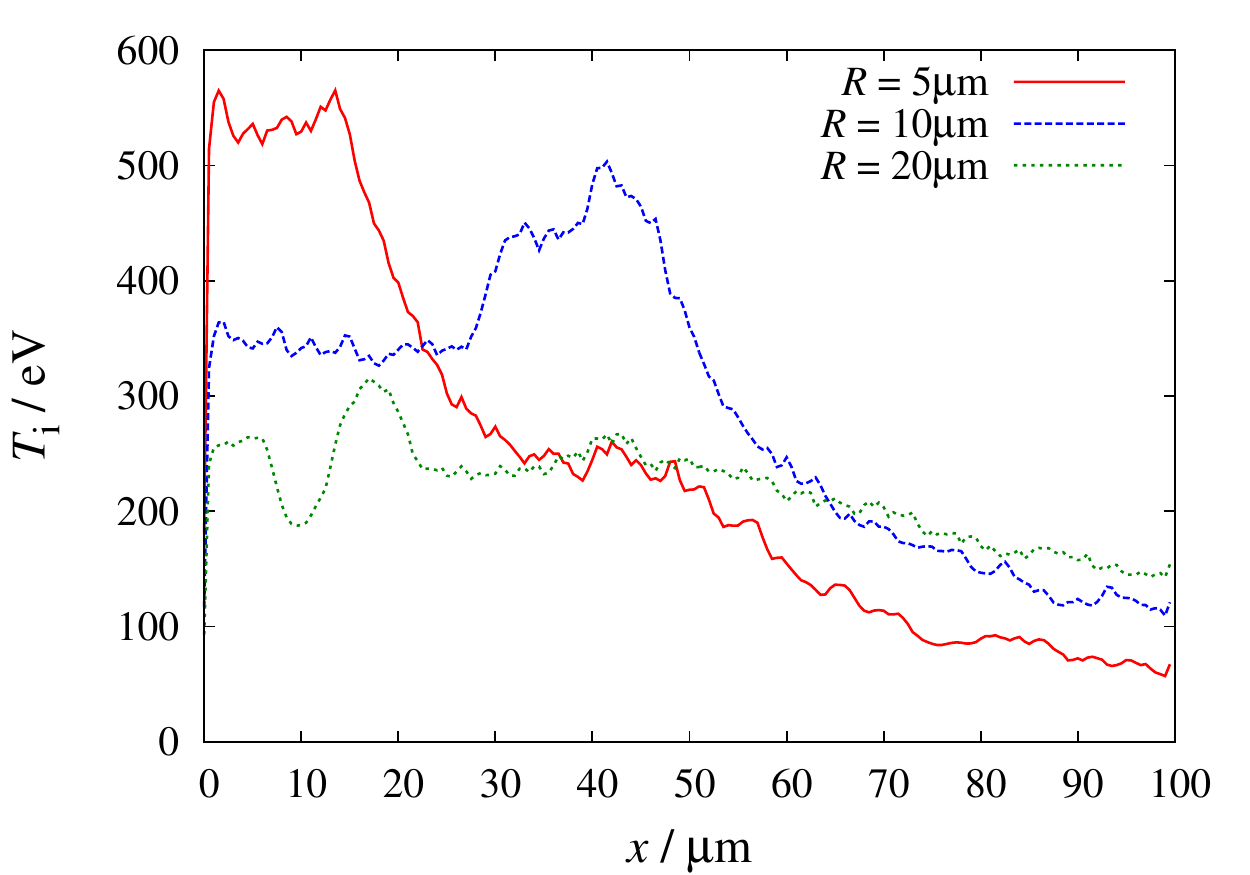}\hfill
\includegraphics[width=0.33\textwidth]{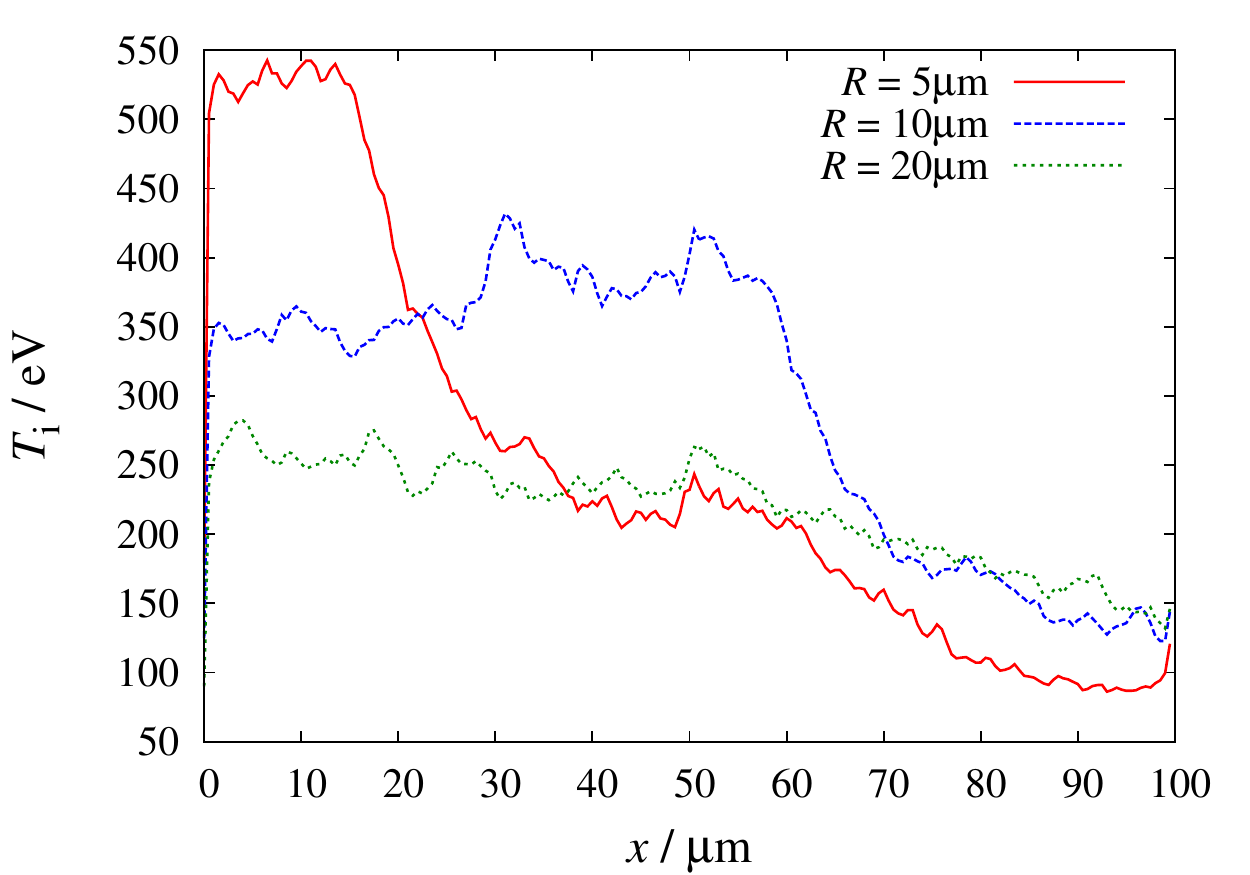}
\end{center}
\caption{The ion temperature at the end of the simulation along the simulation axis for different spot radii $R$. The power is $P=8\times10^{13} W$. The length of the collector is $L=30\mu$m (left panel), $40\mu$m (middle panel), and $50\mu$m (right panel). The opening radius of the collector is twice the laser spot radius. The electron divergence angle is calculated from equation \ref{EqThetaApprox} and the small radius of the collector is chosen to be equal to $R'_{\text{min}}$ given by equation \ref{EqRmin}}
\label{FigTiAxisLowPower}
\end{figure}

Figures \ref{FigTiSurfaceLowPower} and \ref{FigTiAxisLowPower} show the results for runs D$_d$, E$_d$, and F$_d$, with $d=30, 40,$ and $50\mu$m. Here the spot radii are chosen as in runs A$_d$, B$_d$, and C$_d$, but the laser power is reduced by a factor of 4. This means that the laser intensity in runs D$_d$ with $R_{\text{spot}}=5\mu$m is the same as in runs A$_d$ with $R_{\text{spot}}=10\mu$m. Similarly, the laser intensity in runs E$_d$ with $R_{\text{spot}}=10\mu$m is the same as in runs B$_d$ with $R_{\text{spot}}=20\mu$m. The temperature distribution resulting from run D$_{40}$ can be seen in the left panel of figure \ref{FigTiSurfaceLowPower}. The figure shares some similarities with those seen for run A$_{40}$ and some with run B$_{40}$. The temperature initially plateaus but only up to a depth of $x\approx 15\mu$m beyond which it drops off rapidly. The heated region converging from the collector edges at the front surface also converges to around $x\approx 15\mu$m. Figure \ref{FigTiAxisLowPower} shows that this behaviour is independent of the collector length $d$.

For run E$_{40}$ with $R_{\text{spot}}=10\mu$m and low power the behaviour is very similar to run B$_d$ with $R_{\text{spot}}=20\mu$m. The temperature is low near the front surface but increases beyond the tip of the collector deep inside the target. The on-axis temperature profiles for runs E$_d$ shown in figure \ref{FigTiAxisLowPower} show that the location and magnitude of the temperature maximum depend on the value of $d$. For the cases considered here, the optimum length of the collector is $d=30\mu$m. In this case the temperature shows a sharp peak at $x=30\mu$m with a temperature above 560eV, compared to the plateau at 350eV. This corresponds to a factor of 1.6 between the maximum and the baseline. Finally, for runs F$_d$ the laser intensity has the lowest value of all the runs considered here. The low intensity results in a small fast electron temperature according to equation \ref{EqWilksScaling}. The fast electrons are only mildly relativistic and susceptible to the filamentation instability which is enhanced by the low divergence half angle of around $20\degree$. This results in a patchy heating profile and a loss of the electron energy near the front surface of the target, as can be seen in the right panel of figure \ref{FigTiAxisLowPower}.

\section{Summary and Conclusions}

We have performed simulations of fast electron transport with the aim of heating a small region deep inside the target and producing a temperature maximum at finite depth. Using self generated magnetic fields at resistivity gradients, the trajectories of laser generated fast electron current can be manipulated. The return current carried by the background electrons neutralises the fast electron current and causes Ohmic heating. Using this mechanism it is possible to control the heating profile inside the target. It has been shown that principles of geometrical mirror optics can be used to calculate the fast electron trajectories which play the role of the light rays in optics. Using the inventory of non-imaging optics a target geometry was designed similar to a Compound Elliptical Concentrator. We have shown that this geometry allows the energy density deposited to increase at some finite depth inside the target. A factor of 1.6 of the peak ion temperature ratio to the ion temperature at the front surface has been demonstrated with the peak temperature lying roughly $30\mu$m into the target.

The results of the simulations indicate that there is an optimal range for the intensity of the laser beam for focusing fast electrons deep inside the target. Electrons originating from relatively low intensity beams with large spot size and small angular divergence can create a heated region inside the target with a global temperature maximum. At first glance, this seems to be in contradiction to the calculations presented is section \ref{SecEtendue}. These calculations revise the etendue concept of non-imaging optics in the context of fast electron transport. They predict that small sources of fast electrons with large angular spread result in better focusability than large sources with small angular spread resulting from a laser beam with the same total power. The calculations don't take into account any losses of electrons. The conclusion is that the etendue must be seen as a theoretical limit to the focusability of the fast electron beam but cannot be taken as a prediction for the actual heating profile. 

The discrepancy between the simulations and the predictions of section \ref{SecEtendue} is mainly due to electron loss through the magnetic field layers. A small spot size implies a large intensity and this results in a large angular divergence of the fast electrons together with a high average kinetic energy. High energy electrons that impinge on the magnetic field layer at large angles are not effectively reflected due to their increased Larmor radius. These electrons are lost through the boundaries of the collector geometry, reducing the heating further into the target.

\section*{Acknowledgments}
This work is funded by the European Research Council, grant STRUCMAGFAST. The authors would like to thank M.~Tabak for many fruitful discussions.

\bibliographystyle{apsrev}       

\end{document}